\documentclass[letter,twocolumn]{jpsj2} %% for letters
\title{Intrinsic Properties of $\it A$Fe$_{2}$As$_{2}$ ($\it A$ = Ba, Sr) Single Crystal under Highly Hydrostatic Pressure Conditions}

\author{
Kazuyuki \textsc{Matsubayashi}$^{1,3}$$\thanks{E-mail address: kazuyuki@issp.u-tokyo.ac.jp}$,
Naoyuki \textsc{Katayama}$^{1}$$\thanks{Present address: Department of Physics, University of Virginia, Charlottesville, Virginia, USA}$,
Kenya \textsc{Ohgushi}$^{1,3}$,
Atsushi \textsc{Yamada}$^{2,3}$,
Kouji \textsc{Munakata}$^{1,3}$,
Takehiko \textsc{Matsumoto}$^{1}$,
and Yoshiya \textsc{Uwatoko}$^{1,3}$
}

\inst{
$^{1}$Institute for Solid State Physics, The University of Tokyo, Kashiwanoha, Kashiwa, Chiba 277-8581\\
$^{2}$Graduate School of Science and Engineering, Saitama University, Saitama 338-8570\\
$^{3}$JST, TRIP, 5 Sanbancho, Chiyoda, Tokyo 102-0075\\
}

\abst{
We measured the electrical resistivity and ac magnetic susceptibility of BaFe$_{2}$As$_{2}$ and SrFe$_{2}$As$_{2}$ single crystals under pressure using a cubic anvil apparatus. For BaFe$_2$As$_2$, the antiferromagnetic (AF) and structural transitions are suppressed with increasing pressure. Unexpectedly, these transitions persist up to 8 GPa, and no signature of a superconducting transition was observed in the pressure range investigated here. On the other hand, the AF and structural transitions of SrFe$_{2}$As$_{2}$ collapse at around the critical pressure $P_{\rm C}$ $\sim$ 5 GPa, resulting in the appearance of bulk superconductivity. The superconducting volume fraction abruptly increases above $P_{\rm C}$, and shows a dome centered at approximately 6 GPa. Our results suggest that the bulk superconducting phase competes with the AF/orthorhombic phase and only appears in the narrow pressure region of the tetragonal phase.
}

\kword{BaFe$_{2}$As$_{2}$, SrFe$_{2}$As$_{2}$, high pressure, superconductivity, cubic anvil apparatus}

\begin{document}
\maketitle
The recent discoveries of superconductivity in iron-pnictide compounds have attracted much attention in the field of condensed matter physics \cite{Kamihara,Rotter2}. The parent compounds containing FeAs layers exhibit structural and magnetic phase transitions associated with Fe moments. For instance, BaFe$_{2}$As$_{2}$ undergoes structural (tetragonal to orthorhombic) and antiferromagnetic (AF) transitions simultaneously at $T_{\rm s}$ $\sim$ 140 K \cite{Rotter1}. The chemical substitution of Ba for K, Fe for Co and pressure application suppress the AF transition, resulting in the appearance of superconductivity (SC) \cite{Rotter2, Sefat, Alireza}. Similar features are also observed in $A$Fe$_{2}$As$_{2}$ ($A$ = Ca, Sr and Eu) compounds with the same ThCr$_{2}$Si$_{2}$ structure \cite{Torikachvili, Park, Yu, Fukazawa, Igawa, Kotegawa1, Miclea}. Since the emergence of superconductivity coincides with the disappearance of the AF transition, the spin fluctuation of Fe moments is suggested to play an important role in establishing the superconducting ground state.

An important clue to the mechanism of superconductivity should be provided by high-pressure experiment on a stoichiometric sample since the application of pressure introduces no disorder. However, fundamental problems remain to be solved, as the appearance of pressure-induced superconductivity in $A$Fe$_{2}$As$_{2}$ is highly sensitive to pressure homogeneity. In particular for CaFe$_{2}$As$_{2}$, there exists a crucial difference in the presence/absence of superconductivity depending on the hydrostaticity of pressure \cite{Torikachvili, Park, Yu}. The inclusion of only a small amount of tetragonal phase gives rise to a spurious superconductivity in the magnetic and orthorhombic phases. In other words, non hydrostatic pressure may smear out intrinsic properties. Similar discrepancies exist for the pressure effect in BaFe$_{2}$As$_{2}$ and SrFe$_{2}$As$_{2}$. For BaFe$_{2}$As$_{2}$, a single-crystalline sample becomes superconducting above a critical pressure of 2.5 GPa \cite{Alireza}, while a polycrystalline sample exhibits no zero resistance up to 13 GPa under hydrostatic condition using a cubic anvil apparatus \cite{Fukazawa}. On the other hand, the pressure-induced superconductivity in SrFe$_{2}$As$_{2}$ is confirmed by some groups using different high-pressure apparatus; however, the critical pressure for the appearance of SC is controversial \cite{Alireza, Igawa, Kotegawa1}. Furthermore, there is no bulk evidence of the SC transition in BaFe$_{2}$As$_{2}$ and SrFe$_{2}$As$_{2}$, because most previous studies are carried out by resistivity measurement. To resolve these problems, we measured the electrical resistivity and ac magnetic susceptibility of BaFe$_{2}$As$_{2}$ and SrFe$_{2}$As$_{2}$ single crystals under highly hydrostatic pressure conditions up to 8 GPa. In this letter, we report the phase diagrams of these compounds, and present bulk evidence of the pressure-induced superconductivity in SrFe$_{2}$As$_{2}$.

Single crystals were grown by an FeAs self-flux method to avoid the contamination of other elements into the crystals. The starting materials were put in an alumina crucible and sealed in a double quartz tube. The tube was heated up to 1100$^\circ$C, and slowly cooled down to 900$^\circ$C in 50 h. High pressure was generated using a cubic anvil high-pressure apparatus consisting of six tungsten carbide anvils, which has been proved to produce a homogeneous pressure \cite{Mori}. The pressure of the sample is calibrated by the measurement of the resistivity changes of Bi and Te associated with their structural phase transitions at room temperature. The force applied to the sample is kept constant during the measurement by cooling and warming runs. We use glycerin and pyrophyllite as the pressure-transmitting medium and gasket, respectively. Electrical resistivity was measured by a standard four-probe dc technique with current flow in the $ab$ plane. Ac magnetic susceptibility was measured using a conventional Hartshorn bridge circuit at a fixed frequency of 307 Hz. A modulation field with an amplitude of 2 Oe was applied along the $ab$ plane. A similar-size piece of lead was also placed inside the compensated pick-up coil to estimate the magnitude of the signal corresponding to 100$\%$ of the shielding effect to the sample.

\begin{figure}[t]
\begin{center}
\includegraphics[width=0.45\textwidth]{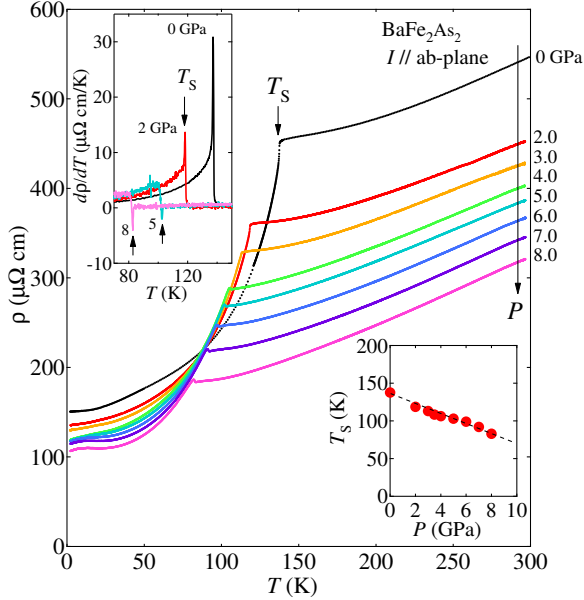}
\end{center}
\caption{(Color online) Temperature dependence of the electrical resistivity of BaFe$_{2}$As$_{2}$ single crystal under pressure. The arrows at $T_{\rm S}$ indicate the location of the AF and structural transition temperatures. Left inset: $d\rho$/$dT$ versus $T$ at selected pressures. Right inset: Pressure dependence of $T_{\rm S}$ obtained by the electrical resistivity measurements. Broken lines are guides for the eye.}
\label{}
\end{figure}

Figure 1 shows the temperature dependence of the electrical resistivity $\rho(T)$ of BaFe$_{2}$As$_{2}$ under pressure. At ambient pressure, $\rho(T)$ decreases with decreasing temperature, and shows a sharp resistivity drop at $T_{\rm S}$ $\sim$ 138 K corresponding to the AF and structural transitions as mentioned above. Here, we define $T_{\rm S}$ as the peak in the derivative $d\rho(T)$/$dT$ (see the left inset of Fig. 1). With increasing pressure, $T_{\rm S}$ shifts to a lower temperature. Above 4 GPa, one may notice that the sharp downward anomaly at $T_{\rm S}$ changes into a slight upturn leading to a peak before decreasing with further decrease in temperature, implying a superzone gap opening. A similar feature is also observed in SrFe$_{2}$As$_{2}$ single crystal under pressure, as shown in Fig. 2. Note that the anomaly at $T_{\rm S}$ on BaFe$_{2}$As remains quite sharp up to the highest pressures, confirming the good hydrostaticity of the pressure environment in the present experiment. The most striking feature of our experiment is that the AF and structural transitions persist against pressure up to 8 GPa, and that there is no signature of a superconducting transition. These results are in contrast to those of an earlier high-pressure study using single-crystalline samples grown by the same method \cite{Alireza}. Here we consider the difference in high-pressure experimental conditions: First, our high-pressure experiments have been performed using a cubic anvil apparatus known to generate hydrostatic conditions owing to the multiple anvil geometry, while the experiment in the previous report was carried out using a diamond anvil cell (DAC) in the uniaxial geometry. Second, the pressure-transmitting medium used in the DAC measurement was Daphne7373, which solidifies at 2.2 GPa at room temperature. The solidification of the liquid pressure-transmitting medium causes inhomogeneous pressure distributions, especially for DAC, which induces uniaxial stress. On the other hand, the glycerin used in our experiments remains near hydrostatic pressure of up to 7 GPa \cite{Osakabe}. Consequently, we speculate that these differences in the high-pressure experimental conditions have a critical influence on the appearance of superconductivity in BaFe$_{2}$As$_{2}$. 

As shown in the right inset of Fig. 1, $T_{\rm S}$ monotonically decreases with decreasing temperature, reaching $\sim$ 84 K at 8 GPa. The slope of $d\rho(T)$/$dT$ $\sim$ -7.0 K/GPa is the smallest among $A$Fe$_{2}$As$_{2}$ compounds ($A$ = Ba, Sr, Eu, Ca), which is consistent with theoretical calculations \cite{Kasinathan, Xie}. We conjecture that the collapse of the structural/magnetic transition may occur above 10 GPa. This deserves further investigation with extension of the pressure range for exploring pressure-induced superconductivity under highly hydrostatic pressure condition.

\begin{figure}[t]
\begin{center}
\includegraphics[width=0.5\textwidth]{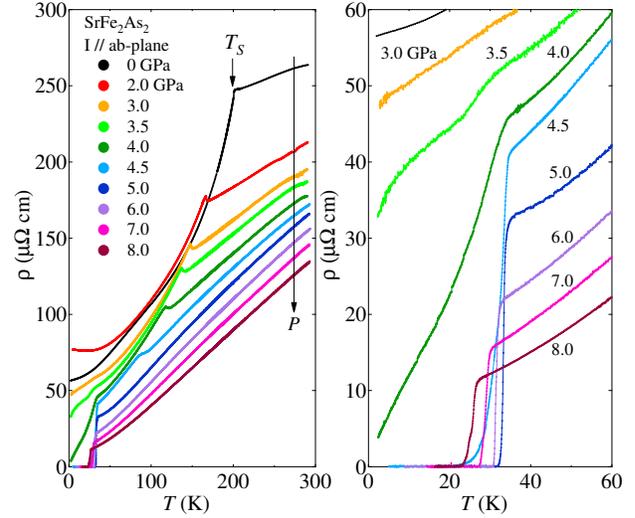}
\end{center}
\caption{(Color online) Temperature dependence of the electrical resistivity of SrFe$_{2}$As$_{2}$ single crystal under pressure. The vertical arrow at $T_{\rm S}$ indicates the location of the AF and structural transition temperatures. The right panel shows the low-temperature part below 60 K.}
\label{}
\end{figure}

Figure 2 shows the electrical resistivity of SrFe$_{2}$As$_{2}$ as a function of temperature for different pressures. At ambient pressure, there is a resistivity anomaly at $T_{\rm S}$ $\sim$ 200 K, corresponding to first-order AF and structural transitions, in agreement with previous reports \cite{Yan, Kumar}. $T_{\rm S}$ monotonically decreases with increasing pressure. Then it seems to collapse at a critical pressure $P_{\rm C}$ $\sim$ 5.0 GPa. In the right panel of Fig. 2, we focus on the low-temperature part of electrical resistivity at selected pressures. At 3.5 and 4.0 GPa, we find a downturn at approximately 30 K. A well-defined transition to a zero-resistance state emerges at 4.5 GPa, where the AF resistive anomaly still exists at $\sim$90 K. At 5.0 GPa, the SC transition sharpens at a transition temperature $T_{\rm SC}$ $\sim$ 32 K. Here, the superconducting transition temperature $T_{\rm SC}$ is defined as a zero resistive temperature. With further increase in pressure, $T_{\rm SC}$ monotonically decreases, and then the superconducting transition tends to broaden again. 

\begin{figure}[t]
\begin{center}
\includegraphics[width=0.5\textwidth]{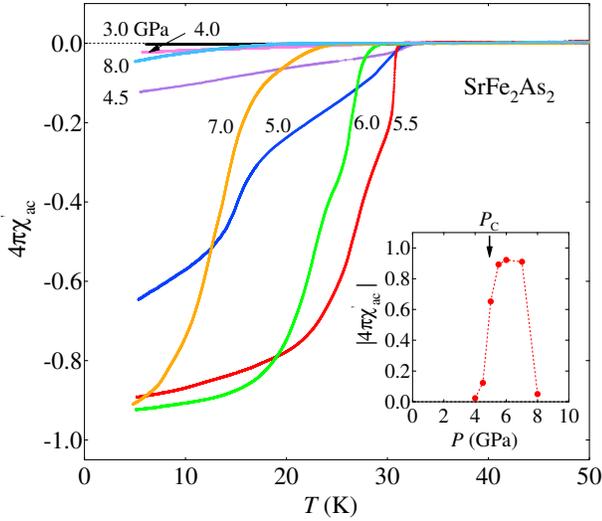}
\end{center}
\caption{(Color online) Temperature dependence of the real part of ac magnetic susceptibility $\chi_{\rm ac}$ for SrFe$_{2}$As$_{2}$ single crystal under pressure, relative to that in normal state. A modulation field of 2 Oe was applied along the $ab$ plane. The inset shows the pressure dependence of the superconducting volume fraction $|4\pi\chi_{\rm ac}'|$ taken at approximately 6 K.}
\label{}
\end{figure}

To establish the bulk nature of the superconducting transition, we carried out ac magnetic susceptibility measurements under pressure. Figure 3 shows the temperature dependence of the real part of $\chi_{\rm ac}$ under pressure, relative to that in the normal state.  At 3.0 GPa, there is no apparent change in $\chi_{\rm ac}'$. With increasing pressure, a noticeable drop appears close to the temperature $T_{\rm SC}$ obtained by the aforementioned resistivity measurements, although the SC diamagnetism at 5.0 GPa is approximately 60$\%$ with a large transition width, implying the presence of a normal state portion in the sample. When pressure is increased up to 5.5 GPa, the transition becomes sharper, and a nearly perfect shielding is detected. This result indicates that the SC transition is of bulk origin. Note that a steplike feature in $\chi_{\rm ac}'$ is observed in the pressure range between 5.0 and 6.0 GPa. This is probably due to the inhomogeneity of the pressure distribution; however, we cannot rule out other possibilities, such as the vortex-glass state causing a double-step superconducting transition \cite{Koziol}. These possibilities need further investigation. Interestingly, the transition begins to broaden again at higher pressures, and thus the shielding effect becomes abruptly weak at 8.0 GPa. To show the pressure dependence of the SC volume fraction, we plot the magnitude of $4\pi\chi_{\rm ac}'$ at the lowest temperature studied here (see the inset of Fig. 3). $|4\pi\chi_{\rm ac}'|$ markedly increases above 5.0 GPa, exhibiting a value ($\sim$1) corresponding to the full shielding effect in a narrow pressure region. Note that this pressure almost coincides with the $P_{\rm C}$ where the AF and structural transitions disappear.

\begin{figure}[t]
\begin{center}
\includegraphics[width=0.47\textwidth]{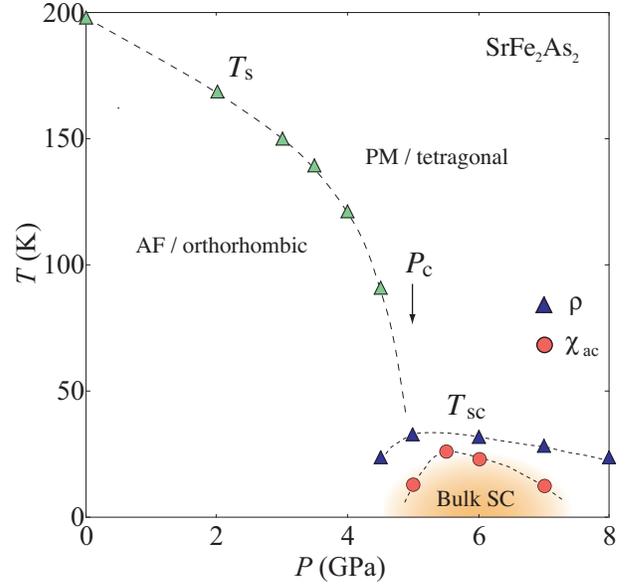}
\end{center}
\caption{(Color online) Temperature-pressure phase diagram of SrFe$_{2}$As$_{2}$. Filled triangles and circles are determined from resistivity and ac susceptibility, respectively. The SC transition temperature $T_{\rm SC}$ is defined as the temperature of zero resistance and 50$\%$ of the full shielding effect. Broken lines are guides for the eye.}
\label{}
\end{figure}

In Fig. 4, we summarize the $T_{\rm S}$ and $T_{\rm SC}$ obtained from our high-pressure experiments to construct a temperature-pressure phase diagram of SrFe$_{2}$As$_{2}$. Here, the SC transition temperature of the ac susceptibility measurement is defined as the temperature at which the sample was observed to reach 50$\%$ of the full shielding effect in the  $\chi_{\rm ac}'$. As the external pressure increases, $T_{\rm S}$ starts to decrease steeply above 3.5 GPa, and then seems to be suppressed to zero in the vicinity of $P_{\rm C}$ $\sim$ 5.0 GPa. The present critical pressure is higher than the $P_{\rm C}$ $\sim$ 3.6 GPa reported by Kotegawa ${\it et}$ ${\it al}$ \cite{Kotegawa1}. The discrepancy can be ascribed to the difference in pressure homogeneity, as observed in the case of BaFe$_{2}$As$_{2}$; the pressure transmitting medium used in previous measurements was Daphne7373. More recently, Kotegawa ${\it et}$ ${\it al.}$ have investigated the pressure-transmitting medium dependence of the pressure-temperature phase diagram by measuring electrical resistivity \cite{Kotegawa2}. They revealed that $P_{\rm C}$ is affected by uniaxial stress; $P_{\rm C}$ was estimated to be 4.4 GPa under better hydrostatic condition. This result is consistent with our experimental observation and supports our good hydrostaticity of the pressure condition.

We return to the features of superconductivity. At 4.5 GPa, we observed the resistivity anomaly due to both antiferromagnetic and superconducting (zero resistance) transitions, implying the coexistence of AF and SC. Indeed, the onset temperature of the shielding effect approximately corresponds to the zero-resistance temperature, but the superconducting volume fraction is quite small. Since the zero resistance due to SC transition can occur even at a small volume fraction, the inhomogeneity of the pressure distribution leads to a deviation from the ideal phase diagram. Furthermore, internal strain induces filamentary superconductivity with zero resistance even at ambient pressure \cite{saha}. To clarify the boundary of the bulk SC, we adopt the $T_{\rm SC}$ determined from ac susceptibility. Our phase diagram indicates that bulk superconductivity only appears above $P_{\rm C}$, where $T_{\rm S}$ is fully suppressed. This is consistent with the rapid sharpening of the superconducting transition width and the remarkable increase in SC volume fraction exceeding $P_{\rm C}$. Note that similar feature has been observed in EuFe$_{2}$As$_{2}$ under pressure, although in this case the magnetically ordered state of the Eu$^{2+}$ moments persists in the SC phase \cite{Terashima}. Consequently, we suggest that the superconductivity does not coexist with the AF/orthorhombic phase. Interestingly, bulk superconductivity is observed only when the pressure is near $P_{\rm C}$ in the paramagnetic tetragonal phase. From this feature, we conjecture that the lattice/magnetic instability in the vicinity of $P_{\rm C}$ plays a crucial role in the appearance of the superconductivity. According to recent NMR measurements for BaFe$_{2}$As$_{2}$ and SrFe$_{2}$As$_{2}$ at ambient pressure, the development of anisotropic spin fluctuations was observed with decreasing temperature in the tetragonal phase \cite{Kitagawa1, Kitagawa2}. Further investigation is needed to determine how AF fluctuations evolve in the pressure region where bulk SC appears.

In summary, we have performed high pressure experiments on ternary iron arsenide BaFe$_{2}$As$_{2}$ and SrFe$_{2}$As$_{2}$ single crystals under highly hydrostatic pressure conditions. For BaFe$_{2}$As$_{2}$, the AF and structural transitions become suppressed with increasing pressure; however, they persist even up to the highest pressure of 8.0 GPa. No signature of a SC transition was observed. Instead, a pressure-induced SC phase was implied to exist at higher pressures. High pressure experiment above 10 GPa is in progress to verify this point. For SrFe$_{2}$As$_{2}$, the bulk nature of the SC transition was confirmed by the ac susceptibility measurement. The most intriguing feature is that bulk superconductivity does not coexist with the AF/orthorhombic phase and is stabilized in the narrow pressure region of the tetragonal phase.

\section*{Acknowledgments}

We thank J. Zhou, M. Hedo, I. Umehara, and K. Kitagawa for helpful discussions and comments. This work was partially supported by a Grant-in-Aid for  Research (No. 21340092) and Young Scientists (No. 19840017) from the Ministry of Education, Culture, Sports, Science and Technology, Japan.

\end{document}